%% file: main.tex
\PassOptionsToPackage{
  colorlinks=true,
  linkcolor=blue,
  citecolor=blue,
  urlcolor=blue
}{hyperref}
\PassOptionsToPackage{table,dvipsnames,svgnames}{xcolor}

\documentclass[a4paper,UKenglish, autoref, cleveref]{lipics-v2021}
\nolinenumbers

\title{Database Theory in Action: Direct Access to Query Answers}

\author{Jiayin Hu}{UC Santa Cruz, USA}{jhu412@ucsc.edu}{https://orcid.org/0009-0004-7505-8255}{}

\author{Nikolaos Tziavelis}{UC Santa Cruz, USA}{ntziavel@ucsc.edu}{https://orcid.org/0000-0001-8342-2177}{}%

\authorrunning{J. Hu and N. Tziavelis} %

\Copyright{Jiayin Hu and Nikolaos Tziavelis} %



\supplement{}
\supplementdetails[subcategory={Source Code}, cite={}, swhid={}]{Software}{https://github.com/hujiayin/direct_access_conjunctive_query}

\category{Database Theory in Action}

\EventEditors{Balder ten Cate and Maurice Funk}
\EventNoEds{2}
\EventLongTitle{29th International Conference on Database Theory (ICDT 2026)}
\EventShortTitle{ICDT 2026}
\EventAcronym{ICDT}
\EventYear{2026}
\EventDate{March 24--27, 2026}
\EventLocation{Tampere, Finland}
\EventLogo{}
\SeriesVolume{365}
\ArticleNo{27}

\usepackage{xspace}
\usepackage[title]{appendix}
\input{macros.tex}

\begin{document}

\maketitle

\begin{abstract}
Direct access asks for the retrieval of query answers by their ranked position, given a query and a desired order.
While the time complexity of data structures supporting such accesses has been studied in depth,
and efficient algorithms for many queries and common orders are known,
their practical performance has received little attention.
We provide an implementation covering a wide range of queries and orders; it allows us to investigate intriguing practical aspects, including the comparative performance of database systems and the relationship between direct access and its single-access counterpart.
\end{abstract}

\section{Introduction}
\label{sec:intro}

Can we simulate a sorted array of answers to a database query so that we can retrieve the answer at any position efficiently?
This question, framed as the \emph{direct access} problem, has been studied intensively in database theory~\cite{Bagan2008,Bourhis25da,braultbaron13thesis,bringmann22da,capelli24da,carmeli23direct,eldar24direct,Keppeler2020Answering}.
Direct access subsumes many computational tasks routinely considered:
Top-$k$~\cite{Rahul19topk} corresponds to accesses to the first $k$ positions,
ranked enumeration~\cite{deep25ranked,tziavelis25anyk} corresponds to accesses to positions $0 \rightarrow 1 \rightarrow \ldots$,
and counting the total number of answers~\cite{Durand20tutorial,PICHLER13counting} can be achieved through binary search for the last position in the simulated array.
On the practical side, efficient direct access can be used to construct an equi-depth histogram on the query result, grouping the answers into equal-size buckets by the chosen order.
It also allows database users to interact with the full set of answers as if they were materialized, even though the actual operations are carried out on the base tables and auxiliary data structures, which are often far more compact.

A variety of theoretical results have been established regarding the feasibility of efficient direct access, depending on the \emph{query}, the desired \emph{order} on the array of answers, as well as the desired \emph{time bounds}.
Most work has focused on \emph{Conjunctive Queries} (CQs)~\cite{bringmann22da,carmeli23direct}, 
but aggregation~\cite{eldar24direct,Keppeler2020Answering} and negation~\cite{capelli24da} have also been considered.
The orders include \emph{lexicographic orders} (e.g., first by $\texttt{city}$ and then by $\texttt{date}$) and
\emph{sum} orders (e.g., by $\texttt{timestamp1} + \texttt{timestamp2}$).
The yardstick of efficiency is typically quasilinear time (i.e., $\bigO(n \log n)$) in the input size $n$ for preprocessing and logarithmic time for each access, but higher preprocessing times have also been considered~\cite{bringmann22da}.
To give an example of an established result, the join of two relations $R_1(A, B) \bowtie_B R_2(B, C)$ admits direct access with quasilinear preprocessing and logarithmic access for the order $A \rightarrow B \rightarrow C$.
However, the order $A \rightarrow C \rightarrow B$ in a certain sense conflicts with the join order, and asumming a certain hypothesis for the complexity of boolean matrix multiplication,
it cannot be supported with these time bounds~\cite{carmeli23direct}.

Despite considerable progress on establishing upper and lower bounds, the only practical implementation of efficient direct access that we are aware of is that of Carmeli et al.\footnote{Their code is available at \url{https://github.com/TechnionTDK/cq-random-enum}.}~\cite{carmeli22random}.
It is intended mainly for random sampling, and so it establishes an arbitrary lexicographic order that cannot be controlled by the user.
We have developed the first, to our knowledge, implementation that 
achieves quasilinear preprocessing and logarithmic access for \emph{all} CQs and lexicographic or sum orders where these bounds are possible.\footnote{Our code is available at \url{https://github.com/hujiayin/direct_access_conjunctive_query}.}
The orders can be either full or partial, in the sense that only a subset of the variables can be used for ranking.
Additionally, it supports the one-off single-access algorithms~\cite{tziavelis23quantiles} (also referred to as \emph{selection}~\cite{carmeli23direct}), covering all quasilinear time cases for the two order types.
After the user specifies a query and an order, it automatically analyzes whether they fall into a tractable case based on the known theoretical results, and then applies the corresponding algorithm~\cite{carmeli23direct,tziavelis23quantiles}.
Our implementation is not integrated with a database system; thus, it should not be thought of as a final solution, but as an impetus to understand practical aspects of direct-access algorithms.
In this paper, we explore the following questions:
\begin{enumerate}
    \item How can direct access be expressed in SQL?
    \item How do existing database systems handle direct access, and when do the more sophisticated algorithms outperform?
    \item How does direct access compare to single-access algorithms in practice?
\end{enumerate}

\section{Direct Access in SQL}

If we want to use an existing system for direct access,
we need to express it in SQL.
We show two ways to achieve this.
One option is to use the OFFSET and LIMIT clauses (\Cref{sql1}); OFFSET skips the first $k$ answers in the ordered result, and LIMIT restricts the output to a fixed number of tuples.  
While effective for accessing consecutive positions, this approach 
cannot express access to multiple, non-consecutive positions in a single query (e.g., in order to find the three quartiles).
Repeating the query with different OFFSETs introduces redundancy and makes common optimization more difficult.
An alternative that is more appropriate for multiple, non-consecutive accesses (\Cref{sql2}) relies on a Common Table Expression (CTE) and the window function \texttt{ROW\_NUMBER()}.
It creates a new attribute that acts as an index to the ordered result, which enables filtering for arbitrary positions (e.g., k1, k2, k3 in the example).  
In the example, we create a CTE named \texttt{ordered\_result} that contains a new column \texttt{row\_idx}, associating each tuple with its ranked position.
Then, we can access positions \texttt{k1, k2, k3} by filtering \texttt{row\_idx} in the WHERE clause.

\begin{figure}[t]
\begin{minipage}[t]{0.37\linewidth}
\begin{lstlisting}[language=SQL, caption= \texttt{OFFSET/LIMIT}, label={sql1}]
SELECT *
FROM R JOIN S ON R.B=S.B
ORDER BY A,B,C
OFFSET k 
LIMIT 1
\end{lstlisting}
\end{minipage}\hfill
\begin{minipage}[t]{0.58\linewidth}
\begin{lstlisting}[language=SQL, caption=CTE with \texttt{ROW\_NUMBER()}, label={sql2}]
WITH ordered_result AS ( 
  SELECT *,
    ROW_NUMBER() 
        OVER (ORDER BY A,B,C) AS row_idx
  FROM R JOIN S ON R.B=S.B)
SELECT * FROM ordered_result
WHERE row_idx IN (k1, k2, k3)
\end{lstlisting}
\end{minipage}
\end{figure}

\section{PostgreSQL vs Theoretical Algorithms}

We now investigate the execution strategies that PostgreSQL 17.4 (\PSQL) employs for direct access, and compare it to our implementation. We restrict our focus to lexicographic orders.

\begin{figure}[t]
    \centering
    \includegraphics[width=\textwidth]{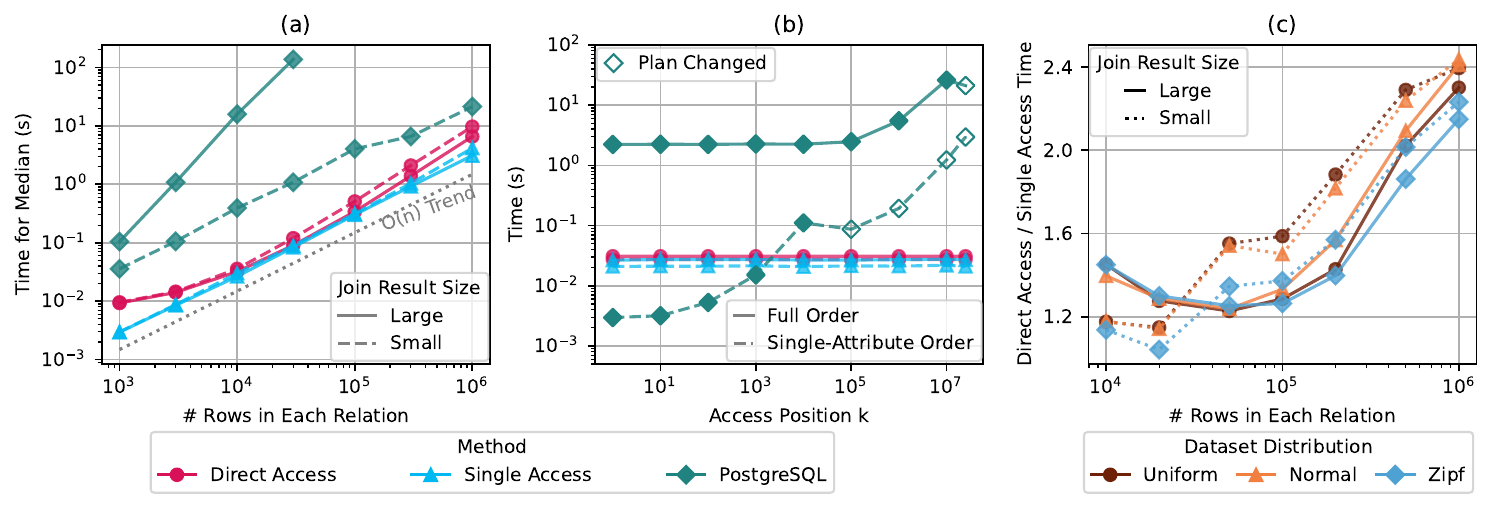} 
    \caption{Experiments on a 3-way join $R(A,B) \bowtie_B S(B,C) \bowtie_C T(C,D)$ using synthetic data. The join result size is controlled by the sampled domain size.
    Left (a): Retrieving the median answer under a full lexicographic order $A \rightarrow B\rightarrow C\rightarrow D$. Middle (b): Time for different $k$ values for relation sizes equal to $10^4$ and a Uniform distribution with a Large join result. Right (c): Ratio of direct access over single access time for accessing the median answer under a full lexicographic order.}
    \label{fig:exp}
\end{figure}

\introparagraph{The standard strategy}
\PSQL typically resorts to the standard strategy of producing and sorting all query answers. 
In general, this strategy is suboptimal;
there are cases where the known theoretical algorithms can achieve quasilinear time, while the number of answers is polynomial.
However, this asymptotic advantage hinges on the worst case of the data distribution, and for a relatively small join result, the standard strategy can be competitive.
\Cref{fig:exp}a illustrates this phenomenon for the median answer of a 3-way join query. 
Under a data distribution that yields a \emph{Large} join result, our direct-access and single-access algorithms outperform \PSQL as the input size (in terms of the number of relation rows) increases.
In contrast, a \emph{Small} join result, whose size is a small constant factor of the input size, gives a trend for \PSQL that is similar to our algorithms and differs only by the constant.

\introparagraph{Optimizations}
Interestingly, we found that \PSQL does employ certain optimizations over the standard strategy in some situations. 
Assuming the position we want to access is $k$, these optimizations attempt to minimize the number of answers that are produced or sorted beyond $k$. 
Since the first $k-1$ answers are never skipped, they can effectively be viewed as optimizations to top-$k$ or ranked enumeration~\cite{deep25ranked,tziavelis25anyk}.
We discuss two of them below.
\begin{itemize}
    \item \introparagraph{Top-N heapsort} This method produces all query answers, but uses a heap to sort only $k$ of them. Compared to other sorting methods, such as quicksort, it is less efficient when $k$ is larger. During query execution, \PSQL decides to switch from Top-N heapsort to quicksort when $k \geq |J| / 2$, where $|J|$ is the number of query answers, or external-memory sort when the working memory is insufficient.
    \item \introparagraph{Sort-before-join} For certain orders, it is possible to insert sorting steps before or between joins, and maintain the ordering at the top of the plan. For example, consider an order on a single attribute $B$ for the join $R(A,B) \bowtie_B S(B,C) \bowtie_C T(C,D)$. One possibility is to perform a sort-merge join between $R$ and $S$, and iterate over this intermediate result in sorted $B$ order for a nested-loop join with $T$. One other possibility is to first join $S$ and $T$, sort the intermediate result on $B$, and then merge-join with $R$. In both cases, we can stop early when $k$ answers are produced.
\end{itemize}

\introparagraph{The effect of $k$}
Our direct-access and single-access algorithms are not sensitive to the value of the accessed position $k$.
In contrast, the performance of \PSQL may heavily depend on that value.
As \Cref{fig:exp}b shows, the running time of the sort-before-join strategy for a single-attribute order outperforms the other algorithms for small $k$, and degrades smoothly with increasing $k$, as we would expect in ranked enumeration.
The figure also shows that the optimizer may decide to change query plan based on $k$.
With increasing $k$ and a single-attribute order, the optimizer switches between the two aforementioned sort-before-join strategies.
For a full lexicographic order where all attributes are ranked and the sort-before-join strategy does not apply,
the optimizer decides to switch from Top-N heapsort to a full sort after $k > 10^7$.

\section{Direct Access vs Single Access}
\label{sec:da_vs_selection}

In a single-access algorithm, no preprocessing is required. 
Each access is computed from scratch, meaning that after one answer is returned, obtaining another requires a new invocation of the algorithm.
The interest in single access lies in its broader applicability; compared to direct access, it can handle a wider class of queries and orders under comparable time bounds.
For instance, quasilinear preprocessing and logarithmic access for free-connex CQs under direct access require restrictions on the lexicographic order, whereas linear-time single access imposes no such limitation~\cite{carmeli23direct}.

When both can be handled efficiently, the natural question is which approach is preferable in practice.
Single access is clearly preferable if only one answer is needed, but how many accesses are required before the additional overhead of direct access pays off?
For lexicographic orders, our implementation achieves quasilinear time for direct access (because of sorting) and linear time for single access (which only uses counting and quickselect).
Thus, we expect the performance gap to widen as the data size increases.
Nevertheless, the asymptotic analysis ignores constant factors and cannot determine the crossover point at which direct access becomes faster.
We use our implementation to shed light on this question.

\Cref{fig:exp}c reports the relative running time of direct access (for one access) compared to single access for a fixed query, a full lexicographic order, and six data distributions.
Because the access cost of direct access is negligible, this ratio can be interpreted as the break-even number of accesses at which direct access becomes preferable.
We generally observe an upwards trend, within $[1,2.4]$ for relation sizes up to $10^6$.
This suggests that even for a small number of accesses (e.g., computing the three quartiles), direct access already offers a performance advantage.

\section{Conclusion}
\label{sec:conclusion}

We have implemented efficient direct-access and single-access algorithms for lexicographic and sum orders.
Our experiments demonstrate promising performance relative to database system strategies, though the latter can be  competitive for small result sizes, simple lexicographic orders, or small $k$ values of accessed positions.
We also found that direct access requires a relatively small number of accesses before it begins to outperform single access in practice.
Looking ahead, further work is needed to integrate these algorithms into database engines, and parallelization offers a clear opportunity to improve their performance.
The archive version of this paper includes more experiments than the ones presented here~\cite{hu2026databasetheoryactiondirect}.

\bibliography{bibliography}

\newpage
\appendix

\section{More Experiments}

\begin{figure}[!htbp]
    \centering
    \includegraphics[width=\linewidth]{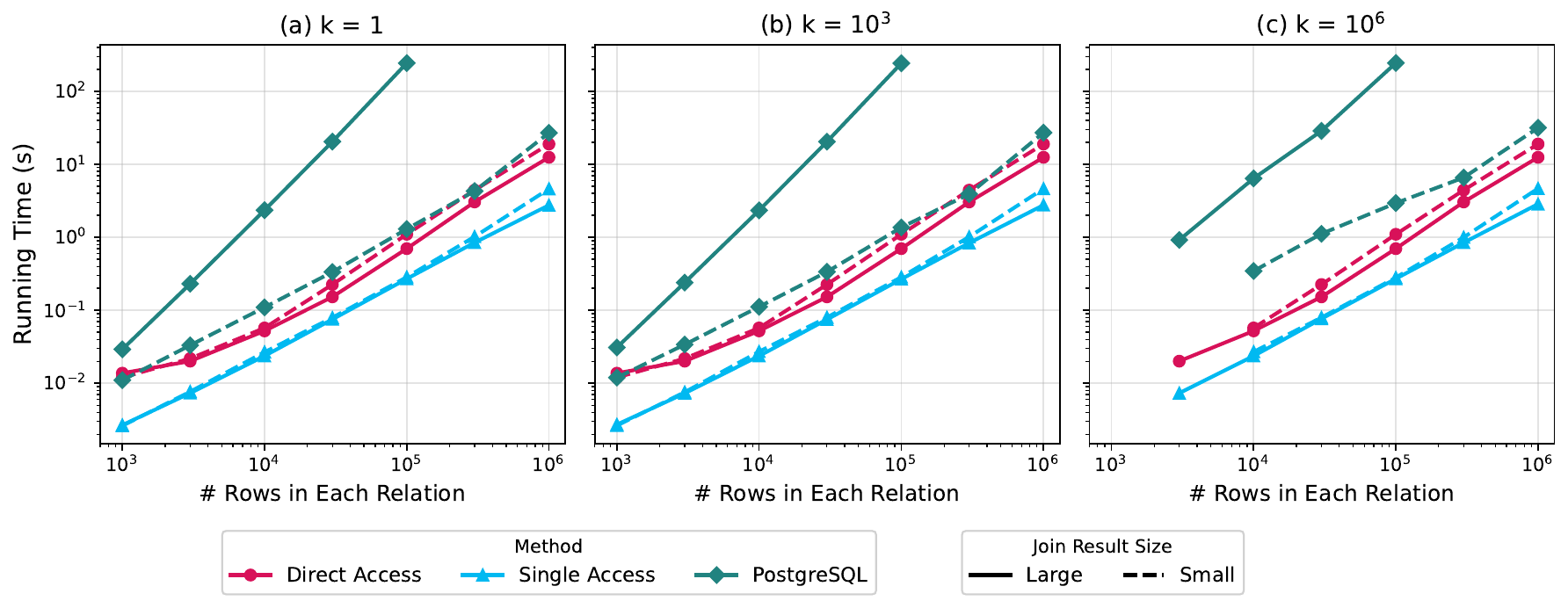}
    \caption{Running time of the 3-way join $R(A,B) \bowtie_B S(B,C) \bowtie_C T(C,D)$ under the full lexicographic order $A \rightarrow B\rightarrow C\rightarrow D$. (a) Retrieving the first answer. (b) Retrieving the answer at position $k=10^3$. (c) Retrieving the answer at position $k=10^6$. Some PostgreSQL results are omitted due to excessive disk consumption. Results are unavailable for a smaller number of rows in each relation in (c) because the total join size is smaller than the target position $k$. }
    \label{fig:fixedK}
\end{figure}

\begin{figure}[!htbp]
    \centering
    \includegraphics[width=1\linewidth]{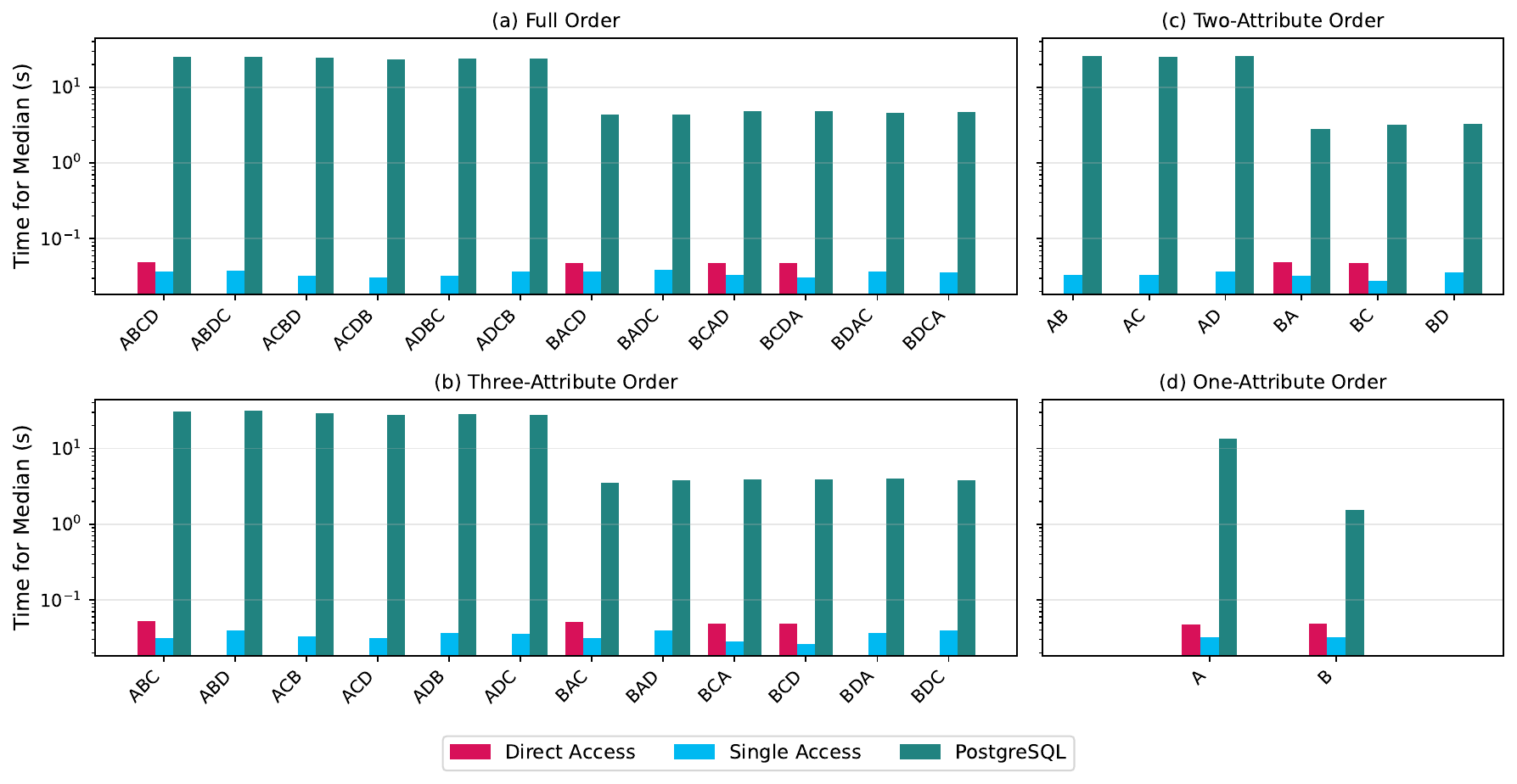}
    \caption{Running time of the 3-way join $R(A,B) \bowtie_B S(B,C) \bowtie_C T(C,D)$ under different lexicographic orders. The x-axis represents different orderings of attributes (e.g., ABC is $A \rightarrow B\rightarrow C$). The subfigures group the experiments by the number of attributes involved in the order: (a) Full Order (four attributes), (b) Three Attributes, (c) Two Attributes, and (d) One Attribute. Direct Access results are only shown for the orders that satisfy the tractability restrictions of the algorithm.}
    \label{fig:orders}
\end{figure}
\vspace*{\fill}

\begin{figure}[t!]
    \centering
    \includegraphics[width=0.75\linewidth]{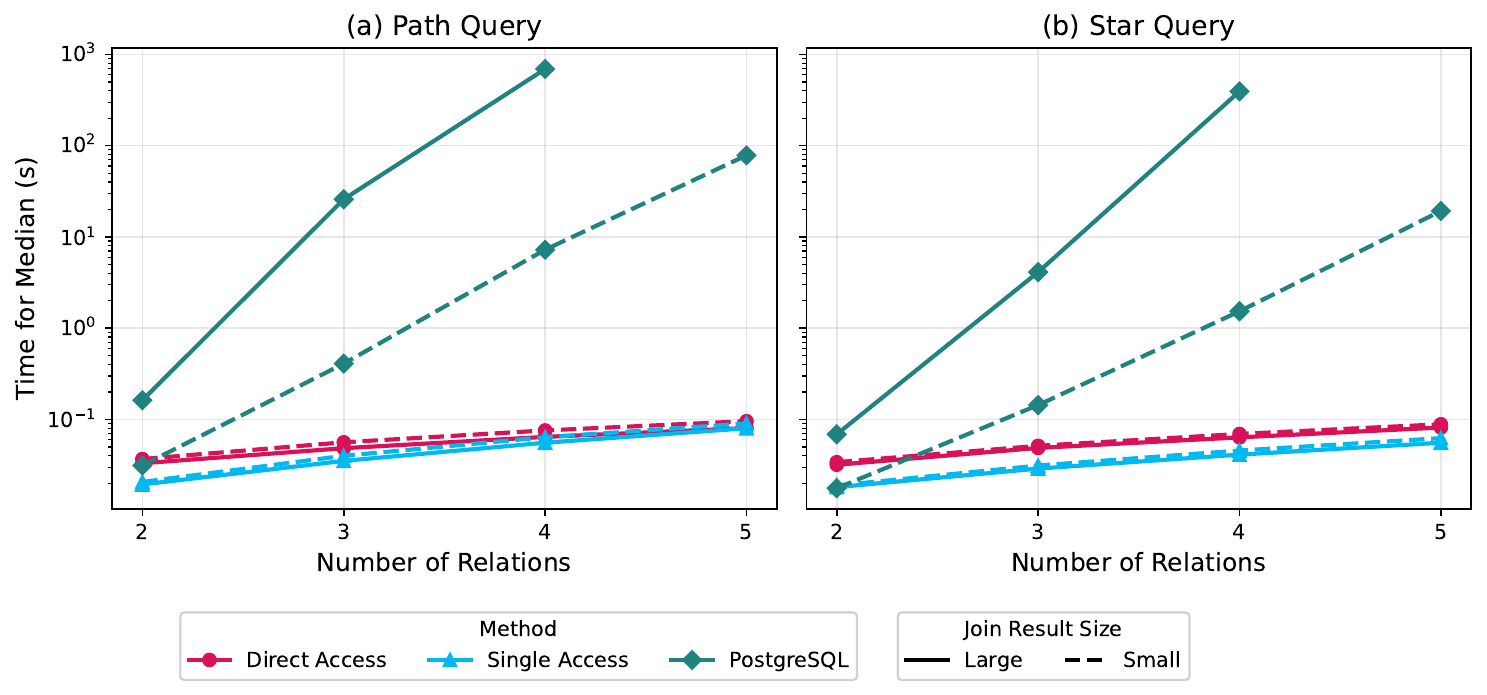}
    \caption{Median execution time for (a) Path Query and (b) Star Query under a full lexicographic order, when increasing the number of relations in the query.}
    
    \label{fig:queries}
\end{figure}
\begin{figure}[t!]
    \centering
    \includegraphics[width=0.4\linewidth]{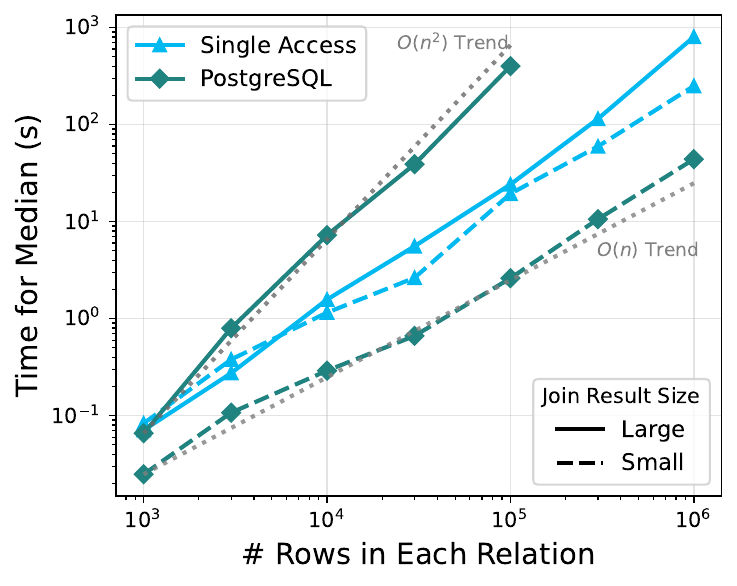}
    \caption{Median execution time of Single Access for 3-way join $R(A,B) \bowtie_B S(B,C) \bowtie_C T(C,D)$ under the sum order A+C. The empirical performance trend of the algorithm is situated between $\bigO(n)$ and $\bigO(n^{2})$. Some PostgreSQL results are omitted due to excessive disk consumption.}
    \label{fig:sumorder}
\end{figure}

\vspace*{\fill}
\clearpage

\end{document}

%% file: macros.tex
\newcommand{\nikos}[1]{{{\color{blue}{[\textbf{Nikos}: #1]}}}}
\newcommand{\jiayin}[1]{{{\color{purple}{[\textbf{Jiayin}: #1]}}}}

\newcommand{\PSQL}{PSQL\xspace}

\newcommand{\bigO}{{\mathcal{O}}}

\newcommand{\introparagraph}[1]{\textbf{#1.}} 

\newcommand{\hide}[1]{} 
\newcommand{\mkclean}{
    \renewcommand{\nikos}{\hide}
    \renewcommand{\jiayin}{\hide}
}

 \mkclean 